\documentclass[a4,11pt]{article}
\usepackage{amssymb}
\usepackage{amsmath}
\usepackage[pdftex]{graphicx}
\usepackage{cite}
\usepackage{slashed}
\usepackage{bm}
\usepackage[top=2.5cm, bottom=2.0cm, left=2.5cm, right=2cm]{geometry}

\begin{document}

\begin{center}
{\large \bf
  The influence of the scalar unparticle on the W - pair production at ILC in the Randall Sundrum model }\\

\vspace*{1cm}

 {\bf Dang Van Soa$^{a,}$\footnote{soadangvan@gmail.com}, Bui Thi Ha Giang$^{b}$\\}

\vspace*{0.5cm}
$^a$ Hanoi Metropolitian University, 98 Duong Quang Ham, Hanoi, Vietnam\\
 $^b$ Hanoi National University of Education, 136 Xuan Thuy, Hanoi, Vietnam
\end{center}

\begin{abstract}
An attempt is made to present the contribution of the scalar unparticle on some scattering processes in the Randall - Sundum (RS) model. We have evaluated the contribution of the scalar unparticle on the W - pair production cross-sections at International Linear Colliders (ILC). The results indicate that 
at the low values of the scaling dimension and the bounds on scale $\Lambda_{U}$ are around few TeV, the cross-sections are much enhanced, which is quite comparable with the W-production in the standard model and hence it is worthwhile to explore in future colliders.\\

\end{abstract}

\section{Introduction}

\hspace*{1cm}The Standard model (SM) is very successful in describing the particle physics. In the Lagrangian of the Standard model, the scale invariance is broken at or above the electroweak scale. 
 At TeV scale, the scale invariant sector has been considered as an effective theory and that if it exists, it is made of unparticle suggested by Geogri \cite{georgi,georgi2}. Based on the Banks-Zaks theory \cite{banks}, unparticle stuff with nontrivial scaling dimension is considered to exist in our world and this opens a window to test the effects of the possible scalar invariant sector, experimentally \cite{chenhe}. Recently, the possibility of the unparticle has been studied with CMS detector at the LHC \cite{cms15, cms16, cms17}. \\
\hspace*{1cm} The effects of unparticle on properties of high energy colliders have been intensively studied in Refs.\cite{zhang,cheung,pra,alan,maj, kuma,sahi,kiku,chen,kha,alie, fried}. In the rest of this work, we restrict ourselves by considering only scalar unparticle. The scalar unparticle propagator is given by
\cite{cheung,georgi2}
\begin{equation}
\Delta_{scalar} = \dfrac{iA_{d_{U}}}{2sin(d_{U}\pi)}(-q^{2})^{d_{U}-2},
\end{equation}
where 
\begin{align}
&A_{d_{U}} = \dfrac{16\pi^{2}\sqrt{\pi}}{(2\pi)^{2d_{U}}}\dfrac{\Gamma\left( d_{U} + \dfrac{1}{2}\right) }{\Gamma(d_{U} - 1)\Gamma (2d_{U})},\\
&(-q^{2})^{d_{U}-2} = \begin{cases}
|q^{2}|^{d_{U} - 2} e^{-d_{U}\pi}   \text{   for s-channel process}, q^{2}  \text{ is positive,}\\
|q^{2}|^{d_{U} - 2}                \text{   for u-, t-channel process}, q^{2}  \text{ is negative.}
\end{cases}
\end{align}
\hspace*{1cm}The effective interactions for the scalar unparticle operators at the scale $\Lambda_{U}$ are given by
\begin{equation}
\lambda_{ff}\dfrac{1}{\Lambda^{d_{U}-1}_{U}}\overline{f}f O_{U}, \lambda_{gg}\dfrac{1}{\Lambda^{d_{U}}_{U}}G_{\alpha\beta}G^{\alpha\beta} O_{U},
\end{equation}
where $d_{U}$ is the scaling dimension of the unparticle $O_{U}$ operators, $\lambda_{i}$ are the dimensionless
effective coupling constants, $G^{\alpha\beta}$ denotes the gauge field strength and $f$ stands for a standard model fermion.\\
\hspace*{1cm} The phenomenology of unparticle physics in models beyond the SM  is discussed\cite{iltan,hao,giang}. In our  previour work\cite{giang}, we have evaluated the contribution of scalar unparticle on the production of Higgs - radion at high energy colliders in the RS model. In this work, we will study the influence of the scalar unparticle on the W - pair production at ILC in the RS model.
 Various ILC physics studies can have a great impact on understanding a new physics around TeV scale. Moreover, it can be transformed into $\gamma\gamma$ collisions with the photon beams generated by 
using the Compton backscattering of the initial electron and laser beams. The photon collider would open a wider window to probe new physics beyond the SM.\\
\hspace*{1cm}The layout of this paper is as follows. In Section II, we give Feynman rules for the couplings of Higgs/radion and scalar unparticle in the RS model. The influence of the scalar unparticle on the W - pair production at ILC is calculated in Section III. Finally, we summarize our results and make conclusions in Section IV.

\section{Feynman rules for the couplings of Higgs/radion and scalar unparticle in the RS model}
\hspace*{1cm}The RS model involves two three-branes bounding a slice of 5D compact anti-de Sitter space. Gravity is localized at the UV brane, while the SM fields are supposed to be localized at the IR brane \cite{rs}. The separation between the two 3-branes leads directly to the existence of an additional scalar called the radion ($\phi$ ), corresponding to the quantum fluctuations of the distance between the two 3-branes. Radion and Higgs boson have the same quantum numbers. General covariance allows a possibility of mixing between the radion and the Higgs boson. The gravity-scalar mixing is described by the following action\cite{csa,dominici,gold}
\begin{equation}
S_{\xi } =\xi \int d^{4}x \sqrt{g_{vis} } R(g_{vis} )\hat{H}^{+} \hat{H},
\end{equation}
where $\xi $ is the mixing parameter, $R(g_{vis})$ is the Ricci scalar for the metric $g_{vis}^{\mu \nu } =\Omega _{b}^{2} (x)(\eta ^{\mu \nu } +\varepsilon h^{\mu \nu } )$ induced on the visible brane, $\Omega _{b} (x) = e^{-kr_{c} \pi} (1 + \frac{\phi_{0}}{\Lambda _{\phi }})$ is the warp factor, $\phi_{0}$ is the canonically normalized massless radion field, $\hat{H}$ is the Higgs field in the 5D context before rescaling to canonical normalization on the brane. 
The mixing of Higgs-radion was given detaily in Refs.\cite{dominici, ahm}. Feynman rules for the couplings of Higgs/radion and the scalar unparticle are showed as follows
\begin{align}
&g_{f\overline{f}h} = i\overline{g}_{f\overline{f}h} = -i\dfrac{gm_{f}}{2m_{W}}\left( d + \gamma b\right),\\
&g_{f\overline{f}\phi} = i\overline{g}_{f\overline{f}\phi} = -i\dfrac{gm_{f}}{2m_{W}}\left( c + \gamma a\right),\\
&\begin{aligned}
g_{WWh} = &i\overline{g}_{Wh}\left[\eta^{\mu\nu} - 2g^{W}_{h}\left(\left(k_{1}k_{2}\right)\eta^{\mu\nu} - k_{1}^{\nu}k_{2}^{\mu}\right)\right]\\
= &igm_{W}\left(d + \gamma b - \gamma b \kappa_{W}\right)\left[\eta^{\mu\nu} - 2g^{W}_{h}\left(\left(k_{1}k_{2}\right)\eta^{\mu\nu} - k_{1}^{\nu}k_{2}^{\mu}\right)\right],
\end{aligned}\\
&\begin{aligned}
g_{WW\phi} = &i\overline{g}_{W\phi}\left[\eta^{\mu\nu} - 2g^{W}_{\phi}\left(\left(k_{1}k_{2}\right)\eta^{\mu\nu} - k_{1}^{\nu}k_{2}^{\mu}\right)\right]\\
= &igm_{W}\left(c + \gamma a - \gamma a \kappa_{W}\right)\left[\eta^{\mu\nu} - 2g^{W}_{\phi}\left(\left(k_{1}k_{2}\right)\eta^{\mu\nu} - k_{1}^{\nu}k_{2}^{\mu}\right)\right],
\end{aligned}
\end{align}
\begin{align}
&\begin{aligned}
g_{\gamma\gamma h} = &iC_{\gamma h}\left[(k_{1}k_{2})\eta^{\mu\nu} - k_{1}^{\nu}k_{2}^{\mu}\right]\\
= &-i\dfrac{\alpha}{2\pi\upsilon_{0}}\left((d + \gamma b)\sum_{i}e_{i}^{2}N_{c}^{i}F_{i}(\tau_{i}) - (b_{2} + b_{Y})\gamma b \right)\\
&\times \left[(k_{1}k_{2})\eta^{\mu\nu} - k_{1}^{\nu}k_{2}^{\mu}\right],
\end{aligned}\\
&\begin{aligned}
g_{\gamma\gamma\phi} =
&iC_{\gamma \phi}\left[(k_{1}k_{2})\eta^{\mu\nu} - k_{1}^{\nu}k_{2}^{\mu}\right]\\
= & -i\dfrac{\alpha}{2\pi\upsilon_{0}}\left((c + \gamma a)\sum_{i}e_{i}^{2}N_{c}^{i}F_{i}(\tau_{i}) - (b_{2} + b_{Y})\gamma a \right)\\
&\times \left[(k_{1}k_{2})\eta^{\mu\nu} - k_{1}^{\nu}k_{2}^{\mu}\right],
\end{aligned}\\
&g_{f\overline{f}U} = i\overline{g}_{f\overline{f}U} = i\dfrac{\lambda_{ff}}{\Lambda_{U}^{d_{U} - 1}},\\
&g_{\gamma\gamma U} = -i\overline{g}_{\gamma\gamma U}\left[(p_{1}p_{2})\eta^{\mu\nu} - p_{1}^{\nu}p_{2}^{\mu}\right] = -4i\dfrac{\lambda_{\gamma\gamma}}{\Lambda_{U}^{d_{U}}} \left[(p_{1}p_{2})\eta^{\mu\nu} - p_{1}^{\nu}p_{2}^{\mu}\right] ,\\
&g_{WWU} = -i\overline{g}_{WW U}\left[(p_{1}p_{2})\eta^{\mu\nu} - p_{1}^{\nu}p_{2}^{\mu}\right] = -4i\dfrac{\lambda_{WW}}{\Lambda_{U}^{d_{U}}} \left[(p_{1}p_{2})\eta^{\mu\nu} - p_{1}^{\nu}p_{2}^{\mu}\right]. 
\end{align}
Here $\gamma = \upsilon /\Lambda _{\phi }, \upsilon = 246$ GeV, $a = -\dfrac{cos\theta}{Z}, b = \dfrac{sin\theta}{Z}, c = sin\theta + \dfrac{6\xi\gamma}{Z}cos\theta, d = cos\theta - \dfrac{6\xi\gamma}{Z}sin\theta$, $\theta$ is the mixing angle, $g^{W}_{h} = \dfrac{\gamma b}{(d + \gamma b - \kappa_{W}\gamma b)m^{2}_{W}}\left(\dfrac{1}{2kb_{0}} + \dfrac{\alpha b_{2}}{8\pi sin^{2}\theta_{W}}\right)$,\\
 $g^{W}_{\phi} = \dfrac{\gamma a}{(c + \gamma a - \kappa_{W}\gamma a)m^{2}_{W}}\left(\dfrac{1}{2kb_{0}} + \dfrac{\alpha b_{2}}{8\pi sin^{2}\theta_{W}}\right)$, $\kappa_{W} = \dfrac{3m^{2}_{W}kb_{0}}{2\Lambda^{2}_{\phi}(k/M_{Pl})^{2}}$, $\dfrac{1}{2}kb_{0} \sim 35$ \cite{ahm}, $b_{3} = 7, b_{2}= 19/6, b_{Y} = - 41/6$, $\theta_{W}$ stands for the Weinberg angle. The auxiliary functions of the $h$ and $\phi$ are given by
\begin{align}
&F_{1/2}(\tau_{i}) = -2 \tau_{i}[1 + (1-\tau_{i}) f(\tau_{i})],\\
&F_1(\tau_{i}) = 2 + 3\tau_{i} + 3\tau_{i}(2-\tau_{i}) f(\tau_{i}),
\end{align}
with
\begin{align}
&f(\tau_{i} ) = \left(\sin ^{-1} \frac{1}{\sqrt{\tau_{i} } } \right)^{2} \, \, \, (for\, \, \, \tau_{i} >1),\\
&f(\tau_{i} ) = -\frac{1}{4} \left(\ln \frac{\eta _{+} }{\eta _{-} } - i\pi \right)^{2} \, \, \, (for\, \, \, \tau_{i} <1),\\
&\eta _{\pm} = 1\pm \sqrt{1 - \tau_{i} } ,\, \, \, \tau _{i} = \left(\frac{2m_{i} }{m_{s} } \right)^{2} .
\end{align}
 Here, $m_{i}$ is the mass of the internal loop particle (including quarks, leptons and W boson), $m_{s}$  is the mass of the scalar state ($h$ or $\phi$), $\tau _{f} = \left(\frac{2m_{f} }{m_{s} } \right)^{2},   \tau _{W} = \left(\frac{2m_{W} }{m_{s} } \right)^{2}$ denote the squares of fermion and W gauge boson mass ratios, respectively.

\section{ The influence of the scalar unparticle on the W - pair production at ILC}
\hspace*{1cm}
  An investigation of W-pair production at ILC plays an important role in testing the SM and searching for physics beyond. In our  previour work\cite{giang}, we have evaluated the contribution of scalar unparticle on the production of Higgs - radion  at high energy colliders in RS model. In this work, we will evaluate the significance 
of the scalar unparticle on the W - pair production at ILC, including the $e^{+}e^{-}$ $\rightarrow$ $W^{+}W^{-}$ process and the $\gamma\gamma$ $\rightarrow$ $W^{+}W^{-}$ subprocess.\\[0.5cm]
\textbf{1. The $\bm{e^{+}e^{-} \rightarrow W^{+}W^{-}}$ collision}\\
\hspace*{1cm}Firstly, we consider the collision process in which the initial state contains electron and positron, the final state contains a pair of $W^{-}$ and $W^{+}$ through the scalar propagators $( \phi , h, U )$, 
\begin{equation} \label{pt4}
e^{-}(p_{1}) + e^{+}(p_{2}) \    \xrightarrow{\phi, h, U} \         W^{-} (k_{1}) + W^{+} (k_{2}).
\end{equation}
The transition amplitude is given by
\begin{equation}
\begin{aligned}
M_{fi} = & i \dfrac{\overline{g}_{ee\phi}\overline{g}_{W\phi}}{q^{2} - m^{2}_{\phi}}\overline{v}(p_{2})u(p_{1}) \varepsilon^{*}_{\mu} (k_{1}) \left[\eta^{\mu\nu} - 2g^{W}_{\phi}\left(\left(k_{1}k_{2}\right)\eta^{\mu\nu} - k_{1}^{\nu}k_{2}^{\mu}\right)\right]\varepsilon^{*}_{\nu} (k_{2})\\
&+ i\dfrac{\overline{g}_{eeh}\overline{g}_{Wh}}{q^{2} - m^{2}_{h}}\overline{v}(p_{2})u(p_{1}) \varepsilon^{*}_{\mu} (k_{1}) \left[\eta^{\mu\nu} - 2g^{W}_{h}\left(\left(k_{1}k_{2}\right)\eta^{\mu\nu} - k_{1}^{\nu}k_{2}^{\mu}\right)\right]\varepsilon^{*}_{\nu} (k_{2})\\
& + i\overline{g}_{eeU}\overline{g}_{WWU} \dfrac{A_{d_{U}}}{2sin(d_{U}\pi)} (-q^{2})^{d_{U} - 2}\overline{v}(p_{2})u(p_{1})\varepsilon^{*}_{\mu} (k_{1}) \left[\left(k_{1}k_{2}\right)\eta^{\mu\nu} - k_{1}^{\nu}k_{2}^{\mu}\right]\varepsilon^{*}_{\nu} (k_{2}).
\end{aligned}
\end{equation}
Here, $q = p_{1} + p_{2} = k_{1} + k_{2}$, $s = (p_{1} + p_{2})^{2}$ is the square of the collision energy.\\
From the expressions of the differential cross-section \cite{pes}
\begin{equation}
\frac{d\sigma}{d(cos\psi)} = \frac{1}{32 \pi s} \frac{|\overrightarrow{k}_{1}|}{|\overrightarrow{p}_{1}|} |M_{fi}|^{2},
\end{equation}
where $\psi = (\overrightarrow{p}_{1}, \overrightarrow{k}_{1})$ is the scattering angle. The model parameters are chosen as $\lambda_{ff} = \lambda_{WW} = \lambda_{0} = 1$, $ m_{h}$ = 125 GeV, $m_{\phi}$ = 10 GeV\cite{soa1}.
As shown in \cite{giang}, the cross section is flat when $d_{U} > 1.6 $, therefore we choose  the $d_{U}$ as $1 < d_{U} < 1.5$. We give estimates for the cross-sections as follows \\
\hspace*{1cm}i) In Fig.1, the total cross-section is plotted as the function of $P_{e^{-}}, P_{e^{+}}$, which are the polarization coefficients of $e^{-}, e^{+}$ beams, respectively. The parameters are chosen as $\sqrt{s} = 1000$ GeV, $d_{U} = 1.1$, $\Lambda_{U} = 1000$ GeV. The figure indicates that
the total cross-section achieves the minimum value when $P_{e^{-}} = P_{e^{+}} = \pm 1$ and the maximum value when $P_{e^{-}} = 1, P_{e^{+}} = -1$ or $P_{e^{-}} = -1, P_{e^{+}} = 1$. \\
\hspace*{1cm}ii) In Fig.2, the total cross-section is plotted as the function of $d_{U}$ in case of $P_{e^{-}} = 1, P_{e^{+}} = -1$. The parameters are chosen as $\sqrt{s} = 1000$ GeV, $\Lambda_{U} = 1000$ GeV. From the figure we can see that the cross section decreases rapidly as $d_{U}$ increases.\\
\hspace*{1cm}iii) In Fig.3, we evaluate the dependence of the total cross-section on the collision energy $\sqrt{s}$ in case of $P_{e^{-}} = 1, P_{e^{+}} = -1$. The collision energy is chosen in the range of 500 GeV$ \leq \sqrt{s} \leq$ 1000 GeV (ILC). The parameters are chosen as $d_{U} = 1.1$, $\Lambda_{U} = 1000$ GeV. The figure shows that the total cross-section increases rapidly when the collision energy $\sqrt{s}$ increases.\\ 
\hspace*{1cm}iv) In Fig.4, we evaluate the dependence of the total cross-section on the $\Lambda_{U}$ at the fixed collision energy, $\sqrt{s} = 1000$ GeV. The polarization coefficients are chosen as $P_{e^{-}} = 1, P_{e^{+}} = -1$, respectively. In case of the additional scalar unparticle propagator, the cross-section decreases rapidly in the region of 1 TeV $\leq \Lambda_{U} \leq$ 3 TeV.\\
\hspace*{1cm} Some numerical values for the cross-section in the case of $d_{U} = 1.1$, $|P_{e^{-}}| = 80\%$, $|P_{e^{+}}| = 30\% $ \cite{bsung1,Clic1} are given in detail in Table 1. The results  show that the cross-section is about $10^{11}$ times larger than that of the W - pair production without the scalar unparticle propagator under the same conditions, which is quite comparable with the W-production in the SM\cite{bosung2}.\\

\textbf{2. The $\bm{\gamma\gamma \rightarrow W^{-}W^{+} }$ subprocess}\\
\hspace*{1cm}Now we consider the collision process in which the initial state contains the couple of photons, the final state contains the pair of $W^{-}$ and $W^{+}$,
\begin{equation} \label{pt4}
\gamma(p_{1}) + \gamma(p_{2}) \    \xrightarrow{\phi, h, U} \         W^{-} (k_{1}) + W^{+} (k_{2}).
\end{equation}
The transition amplitude is given by
\begin{equation}
\begin{aligned}
M_{fi} = & -i \dfrac{C_{\gamma\phi}\overline{g}_{W\phi}}{q^{2} - m^{2}_{\phi}}\varepsilon_{\mu} (p_{1}) \left[\left(p_{1}p_{2}\right)\eta^{\mu\nu} - p_{1}^{\nu}p_{2}^{\mu}\right]\varepsilon_{\nu} (p_{2}) \varepsilon^{*}_{\rho} (k_{1}) \left[\eta^{\rho\sigma} - 2g^{W}_{\phi}\left(\left(k_{1}k_{2}\right)\eta^{\rho\sigma} - k_{1}^{\sigma}k_{2}^{\rho}\right)\right]\varepsilon^{*}_{\sigma} (k_{2})\\
&- i\dfrac{C_{\gamma h}\overline{g}_{Wh}}{q^{2} - m^{2}_{h}}\varepsilon_{\mu} (p_{1}) \left[\left(p_{1}p_{2}\right)\eta^{\mu\nu} - p_{1}^{\nu}p_{2}^{\mu}\right]\varepsilon_{\nu} (p_{2}) \varepsilon^{*}_{\rho} (k_{1}) \left[\eta^{\rho\sigma} - 2g^{W}_{h}\left(\left(k_{1}k_{2}\right)\eta^{\rho\sigma} - k_{1}^{\sigma}k_{2}^{\rho}\right)\right]\varepsilon^{*}_{\sigma} (k_{2})\\
& + i\overline{g}_{\gamma\gamma U}\overline{g}_{WWU} \dfrac{A_{d_{U}}}{2sin(d_{U}\pi)} (-q^{2})^{d_{U} - 2}\varepsilon_{\mu} (p_{1}) \left[\left(p_{1}p_{2}\right)\eta^{\mu\nu} - p_{1}^{\nu}p_{2}^{\mu}\right]\varepsilon_{\nu} (p_{2})\varepsilon^{*}_{\rho} (k_{1}) \left[\left(k_{1}k_{2}\right)\eta^{\rho\sigma} - k_{1}^{\sigma}k_{2}^{\rho}\right]\varepsilon^{*}_{\sigma} (k_{2}).
\end{aligned}
\end{equation}
The effective cross-section $\sigma(s)$ for the $\gamma \gamma \rightarrow W^{-} W^{+}$ subprocess at the ILC can be calculated as follows
\begin{equation}
\sigma_{sub}(s) = \int_{4m_{W}^{2}/s}^{0.83} dx f_{\gamma/e}(x) \int_{(cos\psi)_{min}}^{(cos\psi)_{max}} d cos\psi \dfrac{d\widehat{\sigma}(\widehat{s})}{d cos\psi},
\end{equation} 
where $x = \widehat{s}/s$ in which $\sqrt{\widehat{s}}$ is center of mass energy of the $\gamma \gamma \rightarrow W^{-} W^{+}$ subprocess, $\sqrt{s}$ is center of mass energy of the ILC, $x_{max} = \dfrac{\zeta}{1 + \zeta}$. The photon distribution function $f_{\gamma/e}$ is given by \cite{ginz}
\begin{equation}
f_{\gamma/e} = \dfrac{1}{D(\zeta)}\left[(1 - x)+\dfrac{1}{1 - x} - \dfrac{4x}{\zeta(1 - x)} + \dfrac{4x^{2}}{\zeta^{2}(1 - x)^{2}} \right],
\end{equation}
where 
\begin{equation}
D(\zeta) = \left(1 - \dfrac{4}{\zeta} - \dfrac{8}{\zeta^{2}}\right)ln(1 + \zeta) + \dfrac{1}{2} + \dfrac{8}{\zeta} - \dfrac{1}{2(1 + \zeta)^{2}}.
\end{equation}
For $\zeta = 4.8$, $x_{max} = 0.83$. We estimate the production cross-sections with the contribution of the scalar unparticle propagator as follows \\
\hspace*{1cm}i) In Fig.5, the total cross-section is plotted as the function of $d_{U}$. The parameters are chosen as $\sqrt{s} = 1000$ GeV, $\Lambda_{U} = 1000$ GeV. From the figure we can see that the cross section decreases rapidly as $d_{U}$ increases.\\
\hspace*{1cm}ii) In Fig.6, we evaluate the dependence of the total cross-section on the collision energy $\sqrt{s}$. The collision energy is chosen in the range of 500 GeV$ \leq \sqrt{s} \leq$ 1000 GeV. The parameters are chosen as $d_{U} = 1.1$, $\Lambda_{U} = 1000$ GeV. The figure shows that the total cross-section increases rapidly when the collision energy $\sqrt{s}$ increases.\\ 
\hspace*{1cm}iii) In Fig.7, we evaluate the dependence of the total cross-section on the $\Lambda_{U}$ at the fixed collision energy $\sqrt{s} = 1000$ GeV. The cross-section decreases rapidly in the region of 1 TeV $\leq \Lambda_{U} \leq$ 3 TeV and gradually in the region of 3 TeV $\leq \Lambda_{U} \leq$ 5 TeV .\\
\hspace*{1cm}Some numerical values for the cross-section are given detaily in Table 2. 
The cross-section with $\phi, h, U$ propagators is about $10^{3}$ times larger than that with SM Higgs propagator, however the cross-section is much smaller than that in the $e^{+}e^{-}$ collision.
With the integrated luminosity of the order of $L = 100 fb^{-1}$ yearly \cite{sonm}, the number of events are given in detail in Table 3, which shows that with the contribution of the scalar unparticle, the W - pair production cross-sections may give the observable values at ILC.\\
Note that the phenomenology of the scalar unparticle was discussed recentlly in Ref.\cite{orl} which
showed that the UnCasimir effect could provide the strongest bounds on some restricted region of the unparticle parameter space 
($d_U $ is very closed to $1$).\\ 
\section{Conclusion}
\hspace*{1cm}In this paper, we have evaluated the influence of the scalar unparticle on the W - pair production cross-sections at ILC in the RS model. Numerical evaluations show that the cross section of the W - pair production depends strongly on  the collision energy $\sqrt{s}$, the scaling dimension $d_{U}$ of the unparticle operator $\mathcal{O}_{U}$ and also the energy scale $\Lambda_{U}$. The results indicate that 
at the low values of the scaling dimension ($d_U $ is very closed to $1$) and the bounds on scale $\Lambda_{U}$ are around few TeV, the cross-sections are much enhanced, which is quite comparable with the W-production in the standard model and hence it is worthwhile to explore in future colliders.\\
\hspace*{1cm}Finally, we emphasize that the W mode is the simplest one to study the effect of the scalar unparticle at ILC, this is  due to W - pair can only be produced through the s - channel in the unparticle case.\\
 Acknowledgements: The work is supported in part by the National Foundation for Science and Technology Development (NAFOSTED) of Vietnam under Grant No. 103.01-2016.44.\\
\newpage
\begin{figure}[!htb] \label{fig:eeww3d}
\begin{center}
\includegraphics[width= 8 cm,height= 5 cm]{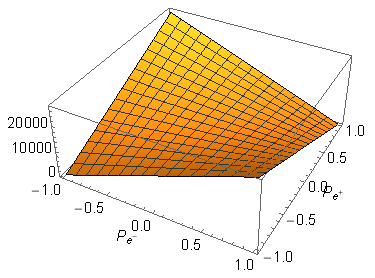}
\caption{ The cross-section as a function of the polarization coefficients ($P_{e^{-}}, P_{e^{+}}$ ) in $e^{+}e^{-} \rightarrow W^{-}W^{+}$ collision.
The parameters are taken to be $\sqrt{s}$ = 1000 GeV, $d_{U}$ = 1.1 and $\Lambda_{U}$ = 1000 GeV.}
\end{center}
\end{figure}
\begin{figure}[!htb] \label{Fig:eewwdu}
\begin{center}
\includegraphics[width= 7.5 cm,height= 4.5 cm]{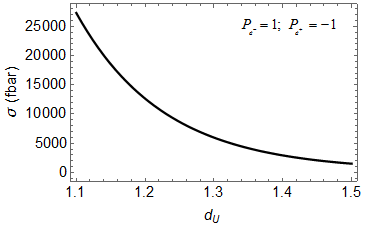}
\caption{The cross-section as a function of the $d_{U}$ in $e^{+}e^{-} \rightarrow W^{-}W^{+}$ collision.
 The parameters are chosen as 
 $P_{e^{-}}$ = 1, $P_{e^{+}} = -1$,
  $\sqrt{s}$ = 1000 GeV and  $\Lambda_{U}$ = 1000 GeV.}
\end{center}
\end{figure}
\begin{figure}[!htb] \label{Fig:eecans}
\begin{center}
\includegraphics[width= 7.5 cm,height= 4.5 cm]{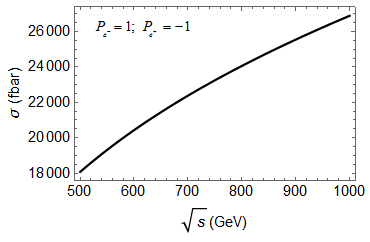}
\caption{ The cross-section as a function of the collision energy $\sqrt{s}$ in $e^{+}e^{-} \rightarrow W^{-}W^{+}$ collision.
The parameters are chosen as $P_{e^{-}}$ = 1, $P_{e^{+}} = -1$, $d_{U}$ = 1.1 and  $\Lambda_{U}$ = 1000 GeV.}
\end{center}
\end{figure}
\begin{figure}[!htb] \label{Fig:eewwlu}
\begin{center}
\includegraphics[width= 7.5 cm,height= 4.5 cm]{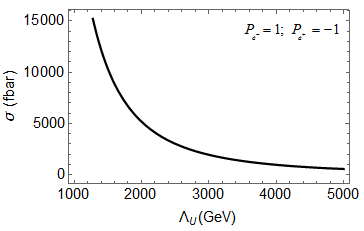}
\caption{The total cross-section as a function of the energy scale $\Lambda_{U}$
in $e^{+}e^{-} \rightarrow W^{-}W^{+}$ collision. 
The parameters are chosen as $P_{e^{-}}$ = 1, $P_{e^{+}} = -1$, $d_{U}$ = 1.1 and $\sqrt{s}$ = 1000 GeV.}
\end{center}
\end{figure}
\begin{figure}[!htb] \label{Fig:ggwwdu}
\begin{center}
\includegraphics[width= 7 cm,height= 4.5 cm]{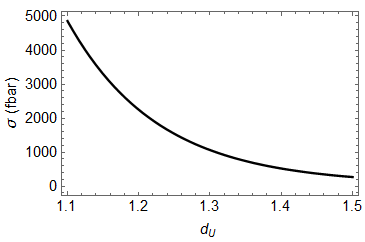}
\caption{The cross-section as a function of the $d_{U}$ in $\gamma\gamma \rightarrow W^{-}W^{+}$ subprocess. The parameters are taken to be $\sqrt{s}$ = 1000 GeV and $\Lambda_{U}$ = 1000 GeV.}
\end{center}
\end{figure}
\begin{figure}[!htb] \label{Fig:ggcans}
\begin{center}
\includegraphics[width= 7.5 cm,height= 4.5 cm]{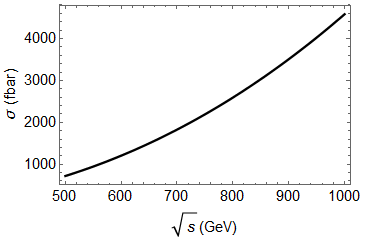}
\caption{The cross-section as a function of the collision energy $\sqrt{s}$ in $\gamma\gamma \rightarrow W^{-}W^{+}$ subprocess. The parameters are chosen as  $d_{U}$ = 1.1 and $\Lambda_{U}$ = 1000 GeV.}
\end{center}
\end{figure}
\begin{figure}[!htb] \label{Fig:ggwwlu}
\begin{center}
\includegraphics[width= 7 cm,height= 4.5 cm]{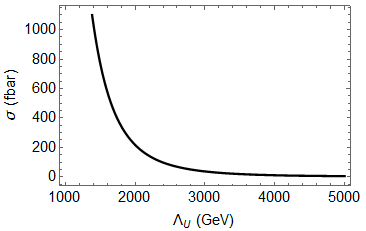}
\caption{The total cross-section as a function of the energy scale $\Lambda_{U}$ in $\gamma\gamma \rightarrow W^{-}W^{+}$ subprocess. The parameters are taken to be  $d_{U}$ = 1.1 and $\sqrt{s}$ = 1000 GeV.}
\end{center}
\end{figure}
\begin{table}[!htb] \label{Fig:eeu}
\begin{tabular}{|c|c|c|c|c|c|c|}
\hline 
$\sqrt{s}$ (GeV) & 500&600&700&800&900&1000 \\ 
\hline 
$\sigma $ ($e^{+}e^{-} \xrightarrow{\phi, h, U} W^{-}W^{+}$) (fbar) & 11384.4& 12871.4&14100.2& 15158.4&16096.0&16943.8\\
\hline
$\sigma $ ($e^{+}e^{-} \xrightarrow{\phi, h} W^{-}W^{+}$) ($10^{-7}$ fbar) & 3.331&3.352&3.366&3.376&3.383&3.388\\
\hline
$\sigma $ ($e^{+}e^{-} \xrightarrow{h_{SM}} W^{-}W^{+}$) ($10^{-7}$ fbar) &3.356&3.378&3.392&3.403&3.410&3.415\\ 
\hline
\end{tabular}
\centering
\caption{Some typical values for the cross-section with the contribution of the scalar unparticle in the $e^{+}e^{-} \rightarrow W^{-}W^{+}$ collisions at the ILC in case of $|P_{e^{-}}| = 80\%  $, $|P_{e^{+}}| = 30\% $ .  The parameters are chosen as $d_{U}$ = 1.1, $m_{h} = 125$ GeV and $m_{\phi} = 10$ GeV.} 
\end{table}

\begin{table}[!htb] \label{Fig:gammau}
\begin{tabular}{|c|c|c|c|c|c|c|}
\hline 
$\sqrt{s}$ (GeV) & 500&600&700&800&900&1000 \\ 
\hline 
$\sigma_{sub} $ ($\gamma \gamma \xrightarrow{\phi, h, U} W^{-}W^{+}$) (fbar) & 720.54&1206.34&1828.06&2593.29&3509.31&4582.99\\
\hline
$\sigma_{sub} $ ($\gamma\gamma \xrightarrow{\phi, h} W^{-}W^{+}$) (fbar) & 0.299&0.442&0.612&0.809&1.032&1.287\\
\hline
${\sigma}_{sub} (\gamma\gamma \xrightarrow{h_{SM}} W^{-}W^{+}) (fbar)$ &0.302&0.446&0.618&0.817&1.042&1.294\\
\hline 
\end{tabular}
\centering
\caption{Some typical values for the cross-section with the contribution of the scalar unparticle in the $\gamma\gamma \rightarrow W^{-}W^{+}$ subprocess at the ILC. The parameters are chosen as in Table 1.}
\end{table}

\begin{table}[!htb] \label{Fig:eventnumber}
\begin{tabular}{|c|c|c|c|c|c|c|}
\hline 
$\sqrt{s}$ (GeV) & 500
&600&700&800&900&1000\\ 
\hline 
N ($\bm{e^{+}e^{-}} \xrightarrow{\phi, h, U} W^{-}W^{+}$) ($10^{6}$) & 1.138&1.287&1.410&1.516&1.609&1.694\\ 
 \hline  
N ($\gamma \gamma \xrightarrow{\phi, h, U} W^{-}W^{+}$) ($10^{5}$) & 0.721&1.206&1.828&2.593&3.509&4.583\\
\hline 
\end{tabular}
\centering
\caption{The number of events in a year with some different values of the collision energy. The parameters are chosen as in Table 1 and the luminosity $L = 100 fb^{-1}$.}
\end{table}

\end{document}